\documentclass[pra, a4paper, preprint, showpacs,superscriptaddress]{revtex4-2}
\usepackage{amsmath,amscd,amsfonts,amssymb, amsbsy, color,bbm, bm,amsthm}
\usepackage{enumitem}
\usepackage{graphicx}{\tiny }
\usepackage{amsbsy}

\theoremstyle{plain}  
\newtheorem{thm}{Theorem}

\newtheorem{cor}{Corollary}
\newtheorem*{pf}{Proof}

\usepackage{longtable}

\numberwithin{equation}{section}
\begin{document}
\title{Computation of sandwiched relative $\alpha$-entropy \\ of two $n$-mode gaussian states}
\author{K. R. Parthasarathy}\email{krp@isid.ac.in}
\affiliation{Indian Statistical Institute, Theoretical Statistics and Mathematics Unit,Delhi Centre,
7 S.~J.~S. Sansanwal Marg, New Delhi 110 016, India} 
\date{\today}
\begin{abstract}	 
A formula for the sandwiched relative $\alpha$-entropy   $\widetilde{D}_\alpha(\rho\vert\vert\sigma)=\frac{1}{\alpha-1}\, \ln\,{\rm Tr}\, \left(\sigma^{\frac{1-\alpha}{2\alpha}}\,\rho\,\sigma^{\frac{1-\alpha}{2\alpha}}\right)^\alpha$ for  $0~<~\alpha~<~1$, of two $n$-mode  gaussian states $\rho$, $\sigma$ in the boson Fock space $\Gamma(\mathbb{C}^n)$ is presented. This computation extensively employs the $\mathcal{E}_2$-parametrization of  gaussian states in $\Gamma(\mathbb{C}^n)$ introduced in J. Math. Phys. {\bf 62}  (2021), 022102.

\vskip 0.5in 

\begin{center}
	{\Large\bf\em To my revered Guru\\Professor C R Rao \vskip 0.2in  on his $\mathbf{101^{\rm st}}$ birthday}
\end{center}

\end{abstract}

\maketitle

\newpage
\section{Introduction}

Sandwiched relative $\alpha$-entropy of two quantum states $\rho$, $\sigma$ was introduced concurrently by Wilde et. al.~\cite{SRE1} and M{\"u}ller Lennert et. al.\cite{SRE2} as   
\begin{eqnarray}
	\widetilde{D}_\alpha(\rho\vert\vert\sigma)=\left\{\begin{array}{l}\frac{1}{\alpha-1}\, \ln\,{\rm Tr}\, \left(\sigma^{\frac{1-\alpha}{2\alpha}}\,\rho\,\sigma^{\frac{1-\alpha}{2\alpha}}\right)^\alpha, {\rm \ if\ } \alpha\in (0,1)\cup (1,\infty), \\ 
		{\rm Tr}\, \rho\left(\ln\rho-\ln\sigma\right), {\rm \ if\ } \alpha=1, \\ 
		\ln\vert\vert\sigma^{-1/2}\,\rho\,\sigma^{-1/2}\vert\vert_\infty, {\rm \ if\ } \alpha=\infty. \end{array} \right.  
\end{eqnarray}
Note that 
\begin{itemize}
\item ${\rm Tr}\, \left(\sigma^{\frac{1-\alpha}{2\alpha}}\,\rho\,\sigma^{\frac{1-\alpha}{2\alpha}}\right)^\alpha=\infty$ if ${\rm supp}(\rho) \nsubseteq \rm{supp}(\sigma)$.
\item  $\lim_{\alpha\rightarrow 1}\, \widetilde{D}_\alpha(\rho\vert\vert\sigma)$ is equal to the quantum relative entropy $D(\rho\vert\vert\sigma)=	{\rm Tr}\, \rho\left(\ln\rho-\ln\sigma\right).$ 
\item $\widetilde{D}_\alpha(\rho\vert\vert\sigma)$ reduces to the Petz--R\'{e}nyi relative entropy~\cite{Renyi1961, Petz1986}
 given by $D_\alpha(\rho\vert\vert\sigma)=\frac{1}{\alpha-1}\ln{\rm Tr}\{\rho^{\alpha}\sigma^{1-\alpha}\},\ \alpha\in (0,1)\cup (1,\infty)$ when  $\rho$ and $\sigma$ commute. Thus sandwiched relative $\alpha$-entropy is viewed  as a non-commutative generalization of  the Petz--R\'{e}nyi relative entropy $D_\alpha(\rho\vert\vert\sigma).$ 
 \item  $\widetilde{D}_\alpha(\rho\vert\vert\sigma)$ reduces to the relative max-entropy~\cite{NDatta2009} 
 $D_{\rm max}=\ln\vert\vert\sigma^{-1/2}\,\rho\,\sigma^{-1/2}\vert\vert_\infty$ in the limit $\alpha\rightarrow \infty$. 
 \item  $\widetilde{D}_\alpha(\rho\vert\vert\sigma)$ is related to the quantum fidelity $F(\rho,\sigma)={\rm Tr}\left(\sigma^{1/2}\rho\sigma^{1/2}\right)^{1/2}$ when $\alpha=1/2$. 
\end{itemize}
Sandwiched relative $\alpha$-entropy $\widetilde{D}_\alpha(\rho\vert\vert\sigma)$ finds several applications in quantum information tasks: It has been employed to prove  strong converse theorems for 
quantum channels~\cite{SRE1,TWW15}; for $\alpha>1$ the sandwiched relative $\alpha$-entropy $\widetilde{D}_\alpha(\rho\vert\vert\sigma)$ has a direct operational interpretation as strong converse error exponent in quantum hypothesis testing~\cite{CMW2014, MTO2015}.  

In this paper we derive a formula for the sandwiched relative $\alpha$-entropy   $\widetilde{D}_\alpha(\rho\vert\vert\sigma)$ for  $0~<~\alpha~<~1$, of two $n$-mode  gaussian states $\rho$, $\sigma$ in the boson Fock space $\Gamma(\mathbb{C}^n)$. We employ the $\mathcal{E}_2$-parametrization of  gaussian states in $\Gamma(\mathbb{C}^n)$ proposed in Ref.~\cite{Tiju-Par-2021} for this computation. 

\section{Mathematical preliminaries}

We begin with the necessary  mathematical preliminaries. All the theorems and proofs that are readily available in previous Refs.~\cite{Par10,Par13,Par-Sen-2015,Tiju-Par-2021,Par21} are only stated.  

 Consider the Hilbert space $L^2(\mathbb{R}^n)$, or equivalently, the  boson Fock space $\Gamma(\mathcal{H})$  over the  complex Hilbert space $\mathcal{H}\equiv\mathbb{C}^n$ of finite dimension $n$.  For any $\mathbf{u}=(u_1,u_2,\ldots , u_n)^T$ in $\mathcal{H}$, define exponential vector $\vert e(\mathbf{u})\rangle$ in the boson Fock space $\Gamma(\mathcal{H})$  by 
\begin{eqnarray*}
\vert e(\mathbf{u})\rangle &=& \sum_{\mathbf{k}\in\mathbb{Z}_+^n}\frac{\mathbf{u}^{\mathbf{k}}}{\sqrt{\mathbf{k}!}}\,\vert \mathbf{k}\rangle =\sum_{0\leq r< \infty}\sum_{\vert\mathbf{k}\vert=r}\, \frac{\mathbf{u}^{\mathbf{k}}}{\sqrt{\mathbf{k}!}}\,\vert \mathbf{k}\rangle  
\end{eqnarray*} 
where $\vert\mathbf{k}\rangle=\vert k_1,k_2,\ldots k_n\rangle$, $\mathbf{k}!=k_1!k_2!\ldots k_n!$ and $\vert \mathbf{k}\vert=k_1+k_2+\ldots k_n.$ 
Then, 
\begin{eqnarray*}
\langle e(\mathbf{u})\vert e(\mathbf{v})\rangle & =& e^{\langle \mathbf{u}\vert \mathbf{v}\rangle}. 
\end{eqnarray*}
The exponential vectors constitute a linearly independent and a total set in  $\Gamma(\mathcal{H})$.  

For any bounded operator $Z$ on $\Gamma(\mathcal{H})$ the generating function  $G_Z(\mathbf{u},\mathbf{v})$,  with $\mathbf{u}, \mathbf{v}$ in $\mathbb{C}^n$,  is defined by ~\cite{Tiju-Par-2021}   
\begin{eqnarray*}
G_Z(\mathbf{u},\mathbf{v})=	\langle e(\bar{\mathbf{u}})\vert \, Z\, \vert e(\mathbf{v})\rangle. 
\end{eqnarray*}
The operator $Z$ is said to belong to the class $\mathcal{E}_2(\mathcal{H})\equiv \mathcal{E}_2$ if 
\begin{equation}
	\langle e(\bar{\mathbf{u}})\vert\, Z\,\vert e(\mathbf{v})\rangle  = c\, \exp \left( \bm{\lambda}^T\mathbf{u}+\bm{\mu}^T\mathbf{v}+\mathbf{u}^T\,A\,\mathbf{u} + \mathbf{u}^T\,\Lambda\,\mathbf{v}+\mathbf{v}^T\,B\,\mathbf{v} \right) ,\ \ \ \forall \ \mathbf{u},\mathbf{v}\in\mathbb{C}^n, 
\end{equation}
where $c\neq 0$ is a scalar; $\bm{\lambda},\ \bm{\mu}\in\mathbb{C}^n$; $A$, $B$ and $\Lambda$ are complex $n\times n$  matrices, with $A,\,B$ being symmetric.  We list the properties~\cite{Tiju-Par-2021} of $Z$ belonging to the operator semigroup  $\mathcal{E}_2$:   
\begin{enumerate}
	\item If $Z\in\mathcal{E}_2$,  then $Z^\dag\in\mathcal{E}_2.$
	\item If $Z_1,\ Z_2\in\mathcal{E}_2$, then $Z_1\, Z_2\in\mathcal{E}_2$. 
	\item The ordered six-tuple $(c,\,\bm{\lambda},\, \bm{\mu},\, A,\,B,\, \Lambda)$ is the  $\mathcal{E}_2$ parametrization of the operator $Z$.	
	\begin{enumerate}
	\item The operator $Z\in \mathcal{E}_2$ is hermitian if and only if  $c$ is real,  $B=\bar{A}$ and  $\Lambda$ is hermitian. 
	\item For any positive operator $Z$ in $\mathcal{E}_2$, its $\mathcal{E}_2$ parameters satisfy  $c>0$, $\bar{\bm{\lambda}}=\bm{\mu}$, $\bar{A}=B$  and $\Lambda\geq 0$.
		\end{enumerate}
\item If  $K$ is a selfadjoint contraction in $\mathcal{H}$, then its second quantization $\Gamma(K)$ is a selfadjoint contraction in $\Gamma(\mathcal{H})$. Furthermore, $\Gamma(K)\in \mathcal{E}_2$ and $\Gamma(K)\,Z\,\Gamma(K)$ denoted by $Z'$ is an element in  $\mathcal{E}_2$ with parameters $(c',\bm{\mu}', A', \Lambda')$ given by $c'=c$, $\bm{\mu}'=K\bm{\mu}$, $A'=KAK^T$, $\Lambda'=K\Lambda\,K$. 
\end{enumerate}

Consider  $Z>0$ with  $\mathcal{E}_2$-parameters  $(c,\,\bm{\mu},\,  A,\, \Lambda)$.  Define a $2n\times 2n$ matrix 
\begin{equation}
M(A,\Lambda)=I_{2n}-\left(\begin{array}{cc} {\rm Re}\Lambda & -{\rm Im}\Lambda \\ {\rm Im}\Lambda & {\rm Re}\Lambda  \end{array}\right)-2\,
\left(\begin{array}{cc} {\rm Re}A & {\rm Im}A \\ {\rm Im}A & -{\rm Re}A  \end{array}\right)
\end{equation}
where $I_{2n}$ denotes $2n\times 2n$ identity matrix. 
If $M(A,\Lambda)\geq 0$ define 
\begin{equation}
c(A,\Lambda)=\sqrt{\det\,M(A,\Lambda)}. 
\end{equation}

\begin{thm} Let $Z$ be a positive  operator in $\mathcal{E}_2$. Then $Z$ is of trace class if and only if $M(A,\Lambda)>0.$ In such a case   
\begin{equation}
\label{trz}
{\rm Tr}\,Z=\frac{c}{c(A,\Lambda)}\, \exp\left[\left(\bm{\mu}_1^T,\,\bm{\mu}_2^T \right)\,M(A,\Lambda)^{-1} \left(\begin{array}{c}
\bm{\mu}_1 \\ \bm{\mu}_2\end{array} \right)  \right], \ \ \bm{\mu}=\bm{\mu}_1+i\, \bm{\mu}_2,\  \bm{\mu}_1,\bm{\mu}_2\in\mathbb{R}^{n}.
\end{equation}
\end{thm}
\begin{pf}
See proof of the Proposition VI.3 of Ref.~\cite{Tiju-Par-2021}. 	 $\hskip 1in\square$
\end{pf}

We parametrize any positive trace-class operator $Z\in \mathcal{E}_2(\mathcal{H})$ by a quadruple of $\mathcal{E}_2$-parameters  $(c,\bm{\mu}, A, \Lambda)$ with $c>0$, $\bm{\mu}\in\mathbb{C}^n$, $A,\, \Lambda\in\mathbb{M}_n(\mathbb{C})$ with $A$ being complex symmetric and $\Lambda$ positive semi-definite.  

\begin{thm} A state $\rho$ in $\Gamma(\mathcal{H})$ is gaussian if and only if $\rho$ belongs to $\mathcal{E}_2(\mathcal{H})$.
\end{thm}
\begin{pf}
	See proof of the Theorem V.7 of Ref.~\cite{Tiju-Par-2021}. 	$\hskip 1in\square$
\end{pf}

\begin{cor}
	If $Z$ is a positive trace class operator in $\mathcal{E}_2(\mathcal{H})$ then $ \frac{Z}{{\rm Tr}Z}$ is a gaussian state. 
\end{cor} 
\begin{pf}
	Follows from the definition of $\mathcal{E}_2(\mathcal{H})$. \end{pf}
 \subsection{Annihilation mean and covariance matrix of a gaussian state}
 
 At every element $\mathbf{u}\in\mathbb{C}^n$ one associates a pair of operators $a(\mathbf{u})$, $a^\dag(\mathbf{u})$,  called annihilation, creation operators~\cite{Tiju-Par-2021,Par10, Par13, Par21}, respectively in the boson Fock space $\Gamma(\mathbb{C}^n)$. There exists a unique unitary operator   
 \begin{equation}
 	W(\mathbf{u})=e^{a^\dag(\mathbf{u})-a(\mathbf{u})}
 \end{equation}
 called the {\em Weyl  operator} on $\Gamma(\mathcal{H})$. 
 With every quantum state $\rho$ in $\Gamma(\mathcal{H})$ we associate a  complex-valued function  
 \begin{equation}
 	\hat{\rho}(\mathbf{u})={\rm Tr}\, W(\mathbf{u})\,\rho,\ \ \  \mathbf{u}\in \mathbb{C}^n
 \end{equation}
called the {\em quantum  characteristic function} of $\rho$ at $\mathbf{u}$. 
\begin{itemize}
\item A quantum state $\rho$ in $\Gamma(\mathcal{H})$  is called a $n$-mode gaussian state if there exists a vector  $\mathbf{m}\in\mathbb{C}^n$, called the {\em annihilation mean} vector, and a real symmetric $2n\times 2n$ matrix $S$ 
such that
\begin{eqnarray}
	\widehat{\rho}\left(\mathbf{u}\right)&=&{\rm exp}\left[-2\,i\, {\rm Im}(\mathbf{x}-i\,\mathbf{y})^T\,\mathbf{m} -  \left(\begin{array}{ll} \mathbf{x}^T, & \mathbf{y}^T\end{array}\right)\, S\, \left(\begin{array}{l} \mathbf{x} \\  \mathbf{y}\end{array}\right) \right] \ \  \nonumber \\
	&=& {\rm exp}\left[-2\,i\, \left(\mathbf{x}^T\,{\rm Im}\,\mathbf{m} -\mathbf{y}^T\,{\rm Re}\,\mathbf{m}\right)- \left(\begin{array}{ll} \mathbf{x}^T, & \mathbf{y}^T\end{array}\right)\, S\, \left(\begin{array}{l} \mathbf{x} \\  \mathbf{y}\end{array}\right) \right]\ \ 
\end{eqnarray}
for all $\mathbf{u}=\mathbf{x}+i\, \mathbf{y}$,  $\mathbf{x},\mathbf{y}\in\mathbb{R}^n.$  
\item Every gaussian state $\rho\equiv\rho(\mathbf{m},S)$ in $\Gamma(\mathbb{C}^n)$ is completely determined by the annihilation mean vector $\mathbf{m}\in\mathbb{C}^n$ and the covariance matrix $S\in\mathbb{M}_{2n}(\mathbb{R})$. 
\end{itemize}
\subsection{Relation between $(\mathbf{m},S)$ and the $\mathcal{E}_2$-parameters of a gaussian state}

The following theorem establishes a connection between $(\mathbf{m},S)$ and the $\mathcal{E}_2$-parameters of a $n$-mode gaussian state.    
 
 \begin{thm}
	Consider a gaussian state  $\rho(\mathbf{m},S)$ with  mean vector $\mathbf{m}\in\mathbb{C}^n$ and $2n\times 2n$ real symmetric covariance matrix $S$. Let the $\mathcal{E}_2$-parameters of $\rho(\mathbf{m},S)$ be $(c,\bm{\mu},A,\Lambda)$. Then 
	\begin{eqnarray}
	c&=&\left[\det\left(\frac{1}{2}I_{2n}+S\right)\right]^{-1/2}\exp\left[\left(\begin{array}{c}{\rm Re}\,\mathbf{m}\\ {\rm Im}\,\mathbf{m}\end{array}\right)^T\,J\, \left(\frac{1}{2}I_{2n}+S\right)^{-1} J\,\left(\begin{array}{c}{\rm Re}\,\mathbf{m}\\ {\rm Im}\,\mathbf{m}\end{array}\right)\right] \\
	\bm{\mu}&=& i\, \left( I_{n}, i\,I_{n}\right)\left(\frac{1}{2}I_{2n}+S\right)^{-1}J\left(\begin{array}{c}{\rm Re}\,\mathbf{m}\\ {\rm Im}\,\mathbf{m}\end{array}\right)\\
	A&=&\frac{1}{4}\,\left( I_{n}, i\,I_{n}\right)\left(\frac{1}{2}I_{2n}+S\right)^{-1}\left(\begin{array}{c}I_{n}\\ i\,I_{n}\end{array}\right) \\ 
	\Lambda&=& I_{n}-\frac{1}{2}\,\left( I_{n}, i\,I_{n}\right)\left(\frac{1}{2}I_{2n}+S\right)^{-1}\left(\begin{array}{c}I_{n}\\ -i\,I_{n}\end{array}\right)
	\end{eqnarray}
where 
\begin{equation}
J=\left(\begin{array}{cc}0 & I_n\\ -I_n & 0\end{array}\right). 
\end{equation}	
In the opposite direction, we have 
\begin{eqnarray}
S&=&M(-A,\Lambda)^{-1}-\frac{1}{2}I_{2n} \\
\left(\begin{array}{c}{\rm Re}\,\mathbf{m}\\ {\rm Im}\,\mathbf{m}\end{array}\right)&=& M(-A,\Lambda)^{-1}\left(\begin{array}{c}{\rm Re}\,\bm{\mu}\\ {\rm Im}\,\bm{\mu}\end{array}\right).
\end{eqnarray}
\end{thm}
     \begin{pf}
     	See proofs of the Propositions VI.1 and VI.3 of Ref.~\cite{Tiju-Par-2021}.  $\hskip 1in\square$
     \end{pf}
\subsection{Gaussian symmetry transformation and structure theorem for $n$-mode gaussian state}
 
 Here we list some important features of gaussian states in $\Gamma(\mathcal{H})$: 
 \begin{itemize}
\item  Any unitary operator  $U\in\mathcal{E}_2(\mathcal{H})$ is a gaussian symmetry i.e., $U\,\rho\,U^\dag$ is a gaussian state whenever $\rho$ is  a gaussian state (see Proposition V.10.1 of Ref.~\cite{Tiju-Par-2021}).  Every gaussian symmetry operation belongs to $\mathcal{E}_2(\mathcal{H})$. 
\item For any gaussian state $\rho(\mathbf{m},S)$ in $\Gamma(\mathbb{C}^n)$  
 there exists a  sequence $0< t_1 \leq t_2 \leq \ldots \leq t_n \leq \infty$ and a symplectic matrix $L\in {\rm Sp}(2n,\mathbb{R})$ such that 
 \begin{equation}
 \label{tLdef}
 \rho(\mathbf{m}, S)=U(\mathbf{m},L)\, \rho(\mathbf{t})  U(\mathbf{m},L)^{-1} 
 \end{equation} 
 where $U(\mathbf{m},L)=W(\mathbf{m})\, \Gamma(L)$ is a  unitary gaussian symmetry operator~\cite{Par21} consisting  of a phase space translation $W(\mathbf{m})$  and a disentangling unitary transformation $\Gamma(L)$  on the boson Fock space $\Gamma(\mathbb{C}^n)$. Here  
 \begin{eqnarray}
 \label{thermal}
 \rho(\mathbf{t})&=& \rho(\mathbf{0},\, D(\mathbf{t})) \nonumber \\ 
 &=&  \rho(t_1)\otimes\rho(t_2)\otimes \ldots \otimes \rho(t_n)  \\
 \rho(t_j)&=&p(t_j)\,\sum_{k_j=0}^{\infty}e^{-k_j\,t_j}\,\vert k_j\rangle\langle\,k_j\vert,\ \ \ p(t_j)=(1-e^{-t_j}),\ j=1,2,\ldots n \nonumber
 \end{eqnarray}
 corresponds to a $n$-mode gaussian thermal state characterized by zero mean and  covariance matrix $ D(\mathbf{t})$ given by 
 \begin{eqnarray}
 \label{SEdef}
 D(\mathbf{t})&=& L^T\, S\, L = \left(\begin{array}{cc} D_0({\mathbf{t}}) &  0 \\ 
 0 & D_0({\mathbf{t}})  \end{array}   \right),  \nonumber \\
 D_0({\mathbf{t}})&=&{\rm diag}\left[\frac{1}{2}\coth\left(\frac{t_j}{2}\right), j=1,2,\ldots n\ \right].
 \end{eqnarray} 
 \end{itemize}  
  Thus every gaussian state in  $\Gamma(\mathbb{C}^n)$ is characterized by three equivalent fundamental parametrizations~\cite{Par10,Par13,Tiju-Par-2021,Par21}: 
 \begin{enumerate}
 	\item   $(\mathbf{m},\, S)$: mean annihilation vector $\mathbf{m}\in\mathbb{C}^n$    and real symmetric covariance matrix $S\in\mathbb{M}_{2n}(\mathbb{R})$.  	
 	\item  $(\mathbf{t},\, L):$  Thermal parameters   $\mathbf{t}=(t_1,t_2,\ldots , t_n), \   0< t_1 \leq t_2 \leq \ldots \leq t_n \leq \infty$ and  $L\in$ Sp(2n,$\mathbb{R}$) such that $\rho(\mathbf{m}, S)=U(\mathbf{m},L)\, \rho(\mathbf{t})  U(\mathbf{m},L)^{-1}$ (see (\ref{tLdef}), (\ref{thermal}), and  (\ref{SEdef})).
 	\item $(c,\bm{\mu},\, A, \Lambda)$: $\mathcal{E}_2(\mathcal{H})$-parameters with $c>0$, $\bm{\mu}\in\mathbb{C}^n$, $A,\, \Lambda\in\mathbb{M}_n(\mathbb{C})$ with a complex symmetric $A$ and positive semi-definite $\Lambda$.
 \end{enumerate} 
 
\section{Computation of sandwiched relative $\alpha$-entropy $\bm{\widetilde{D}_\alpha(\rho\vert\vert\sigma)}$ \\ of  two gaussian states $\bm{\rho,\ \sigma}$}
 
 The $\alpha$-dependent sandwiched R{\'e}nyi relative entropy~\cite{SRE1,SRE2} between two  states $\rho,\ \sigma$ is given by  
 \begin{eqnarray*}
 \widetilde{D}_\alpha(\rho\vert\vert\sigma)=\frac{1}{\alpha-1}\, \ln\,{\rm Tr}\, \left[\left(\sigma^{\frac{1-\alpha}{2\alpha}}\,\rho\,\sigma^{\frac{1-\alpha}{2\alpha}}\right)^\alpha\right]. 
 \end{eqnarray*}
 Let $\sigma$, $\rho$ be two $n$-mode gaussian states with       
 \begin{eqnarray}
 \sigma'&=& U(\bm{\ell}, L)\,  \sigma  \, \left(U(\bm{\ell}, L)\right)^{-1}=\rho(\mathbf{s})  \\ 
 \rho'&=&U(\bm{\ell}, L)\,  \rho  \, \left(U(\bm{\ell}, L)\right)^{-1}
 \end{eqnarray}
 where $\rho(\mathbf{s})$ is a $n$-mode  thermal state   
 \begin{eqnarray}
 \label{rhos}
 \rho(\mathbf{s})&=&\rho(s_1)\otimes\rho(s_2)\otimes \ldots \otimes \rho(s_n), \\
 \rho(s_j)&=&p(s_j)\,\sum_{k_j=0}^{\infty}e^{-k_j\,s_j}\,\vert k_j\rangle\langle\,k_j\vert,\ \ \ p(s_j)=(1-e^{-s_j}),\ j=1,2,\ldots n. \nonumber 
   \end{eqnarray} 
 characterized by the parameters $\mathbf{s}=(s_1,s_2,\ldots , s_n), \   0< s_1 \leq s_2 \leq \ldots \leq s_n \leq \infty$. Note that  $\rho(\infty)=\vert\Omega\rangle\langle \Omega\vert$  denotes the  1-mode  Fock vacuum state and  $p(\infty)=1.$ 
 
   
 The $\alpha$-dependent sandwiched relative entropy remains invariant when both the states $\rho$, $\sigma$ are
   changed by any  unitary transformation $U$. Thus   
 \begin{eqnarray}
 \widetilde{D}_\alpha(\rho\vert\vert\sigma)&=&
 \widetilde{D}_\alpha(\rho'\vert\vert\sigma') \nonumber \\
                                           &=& \widetilde{D}_\alpha(\rho'\vert\vert\rho(\mathbf{s})) \nonumber \\
                                           &=&  \frac{1}{\alpha-1}\,\ln\, {\rm Tr}\,  \left\{\rho(\mathbf{s})^{\frac{1-\alpha}{2\alpha}} \,\rho'\, \rho(\mathbf{s})^{\frac{1-\alpha}{2\alpha}}\right\}^\alpha. 
                                           \end{eqnarray}
Let us denote
\begin{eqnarray}
T_\alpha(\rho',\rho(\mathbf{s}))={\rm Tr}\,  \left\{\rho(\mathbf{s})^{\frac{1-\alpha}{2\alpha}} \,\rho'\, \rho(\mathbf{s})^{\frac{1-\alpha}{2\alpha}}\right\}^\alpha.
\end{eqnarray}
Putting
$$p(\mathbf{s})=\prod_{j=1}^n p(s_j),$$ 
 we obtain from (\ref{rhos}) 
\begin{eqnarray}
\rho(\mathbf{s})^{\frac{1-\alpha}{2\alpha}}&=&p(\mathbf{s})^{\frac{1-\alpha}{2\alpha}}  \, \sum_{\mathbf{k}\in\mathbb{Z}_+^n} e^{-\sum_{j=1}^n\, k_j\,s_j\,\left(\frac{1-\alpha}{2\alpha}\right)}\vert \mathbf{k}\rangle\langle\mathbf{k}\vert  \nonumber \\ 
&=& p(\mathbf{s})^{\frac{1-\alpha}{2\alpha}}  \, \Gamma(K)
\end{eqnarray}
where 
\begin{equation}
K={\rm diag}\left(e^{-s_1\,\left(\frac{1-\alpha}{2\alpha}\right)},e^{-s_2\,\left(\frac{1-\alpha}{2\alpha}\right)},\ldots ,e^{-s_n\,\left(\frac{1-\alpha}{2\alpha}\right)}  \right)
\end{equation}
 is the contraction diagonal matrix and $\Gamma(K)\in\mathcal{E}_2(\mathcal{H})$ is the corresponding positive contraction operator ~\cite{Tiju-Par-2021}.     
Thus 
\begin{eqnarray}
 \left\{\rho(\mathbf{s})^{\frac{1-\alpha}{2\alpha}} \,\rho'\, \rho(\mathbf{s})^{\frac{1-\alpha}{2\alpha}}\right\}^\alpha&=& 
 p(\mathbf{s})^{1-\alpha}\,\left\{\Gamma(K)\rho'\,\Gamma(K) \right\}^\alpha. 
 \end{eqnarray}
Consider the positive trace class operator $Z\in\mathcal{E}_2(\mathcal(H))$ defined by 
\begin{equation}
Z=\Gamma(K)\rho'\,\Gamma(K).
\end{equation}
Suppose the transformed gaussian state $\rho'$ has its $\mathcal{E}_2$-parameters $\left(c,\bm{\mu},A, \Lambda\right).$ It follows that $Z$ is an $\mathcal{E}_2$ operator with parameters $\left(c',\bm{\mu}',A', \Lambda'\right)=\left(c,K\bm{\mu},K\,A\,K^T, K\Lambda\,K\right)$. Then (see (\ref{trz})) 
\begin{equation}
\label{trzf}
{\rm Tr}\,Z=\frac{c}{c(A',\Lambda')}\, \exp\left[\left(\bm{\mu}_{1}^{'\,T},\,\bm{\mu}_{2}^{'\,T} \right)\,M(A',\Lambda')^{-1} \left(\begin{array}{c}
\bm{\mu}'_{1} \\ \bm{\mu}'_{2}\end{array} \right)  \right],  
\end{equation}
where $c(A',\Lambda')=\sqrt{\det\, M(A',\Lambda')}.$

Now  $\rho_Z=\frac{Z}{{\rm Tr}\, Z}$ is a gaussian state  with  $\left(\frac{c}{{\rm Tr}\,Z},\bm{\mu}',A', \Lambda'\right)$   as its $\mathcal{E}_2$-parameters.  From the last part of Theorem~3  the covariance matrix $S_Z$ of $\rho_Z$ is given by  
\begin{equation}
S_Z=M(-A',\Lambda')^{-1}-\frac{1}{2}\,I_{2n}.
\end{equation} 
Through Williamson resolution~\cite{Par10,Par13,Par21} of the covariance matrix  viz.,  
\begin{eqnarray}
\label{SESz}
D(\mathbf{t}_Z)&=& L_Z^T\, S_Z\, L_Z = \left(\begin{array}{cc} D_{0}({\mathbf{t}_Z}) &  0 \\ 
0 & D_{0}({\mathbf{t}_Z})  \end{array}   \right),\ \ L_Z\in {\rm Sp}(2n,\mathbb{R}), \\
D_{0}({\mathbf{t}_Z})&=&{\rm diag}\left[\frac{1}{2}\coth\left(\frac{\left(\mathbf{t}_{Z}\right)_j}{2}\right), j=1,2,\ldots n\ \right].
\end{eqnarray} 
we construct $\rho(\mathbf{t}_Z)$, equivalent to $\rho_Z$ by a unitary gaussian symmetry,  with thermal parameters 
$\mathbf{t}_Z=\left( (t_{Z})_1\leq (t_{Z})_2\leq \ldots (t_{Z})_n\right)$. 

Thus
\begin{eqnarray}
{\rm Tr}\,\rho_Z^\alpha&=&  {\rm Tr}\,\rho(\mathbf{t}_Z)^\alpha \nonumber \\
&=& \frac{p(\mathbf{t}_Z)^\alpha}{p(\alpha\,\mathbf{t}_Z)}. 
\end{eqnarray}
Therefore 
\begin{equation}
{\rm Tr}\, Z^\alpha=\frac{\left[p(\mathbf{t}_Z)\right]^\alpha}{p(\alpha\,\mathbf{t}_Z)}\, \left({\rm Tr}\, Z\right)^\alpha. 
\end{equation}  
The following theorem summarizes the above computations:   

\begin{thm} Let $\rho$, $\sigma\equiv\rho(\mathbf{\ell},S)$ be two $n$-mode gaussian states in $\Gamma(\mathbb{C}^n)$, with $\mathbf{\ell}$, $S$ denoting the annihilation mean and covariance matrix of $\sigma$. Let $U(\bm{\ell}, L)$ be the gaussian symmetry leading to the standard form of  $\sigma$ i.e.,  
	\begin{eqnarray*}
\rho(\mathbf{s})	 &=&U(\bm{\ell}, L)\, \sigma\, \left(U(\bm{\ell}, L)\right)^{-1}\,\,  	 \, = \rho(s_1)\otimes\rho(s_2)\otimes \ldots \otimes \rho(s_n), \\
		\rho(s_j)&=&p(s_j)\, \ \ \sum_{k_j=0}^{\infty}e^{-k_j\,s_j}\,\vert k_j\rangle\langle\,k_j\vert; 
		\ \  \ p(s_j)=(1-e^{-s_j}),\ j=1,2,\ldots n.	
	      		\end{eqnarray*}   
Let  $\left(c,\bm{\mu},A, \Lambda\right)$  be the $\mathcal{E}_2$-parameters of the transformed gaussian state $\rho'=U(\bm{\ell}, L)\, \rho\, \left(U(\bm{\ell}, L)\right)^{-1}$. Consider the   positive trace-class operator $Z\in\mathcal{E}_2(\mathcal{H})$, characterized by its $\mathcal{E}_2$-parameters  $\left(c',\bm{\mu}',A', \Lambda'\right)=\left(c,K\bm{\mu},K\,A\,K^T, K\Lambda\,K\right)$, where   $K={\rm diag}\left(e^{-s_1\,\left(\frac{1-\alpha}{2\alpha}\right)},e^{-s_2\,\left(\frac{1-\alpha}{2\alpha}\right)},\ldots ,e^{-s_n\,\left(\frac{1-\alpha}{2\alpha}\right)}  \right)$ is a selfadjoint contraction in $\mathcal{H}$.   Suppose the gaussian state $\rho_Z=\frac{Z}{{\rm Tr}\, Z}$, constructed from $Z$,  has its thermal parameters  $\mathbf{t}_Z=\left( (t_{Z})_1\leq (t_{Z})_2\leq \ldots (t_{Z})_n\right)$.
Then, the  $\alpha$-dependent sandwiched relative entropy  $\widetilde{D}_\alpha(\rho\vert\vert\sigma)=\frac{1}{\alpha-1}\, \ln\,{\rm Tr}\, \left(\sigma^{\frac{1-\alpha}{2\alpha}}\,\rho\,\sigma^{\frac{1-\alpha}{2\alpha}}\right)^\alpha$, for \  $0~<~\alpha~<~1$, of  $\rho$ and $\sigma$ is given by  
	\begin{eqnarray}
		\widetilde{D}_\alpha(\rho\vert\vert\sigma)&=& \widetilde{D}_\alpha(\rho'\vert\vert \rho(\mathbf{s})) \nonumber \\ 
		& =& \frac{1}{\alpha-1}\,\ln\, T_\alpha(\rho',\rho(\mathbf{s})),  
	\end{eqnarray}
where 
	\begin{eqnarray}
		T_\alpha(\rho',\rho(\mathbf{s}))&=&\frac{p(\mathbf{s})^{1-\alpha}\, p(\mathbf{t}_Z)^\alpha}{p(\alpha\,\mathbf{t}_Z)}\, \left({\rm Tr}\, Z\right)^\alpha,  \\ 
		p(\mathbf{s})&=&\prod_{j=1}^n p(s_j), \ \ 	p(\mathbf{t}_Z)=\prod_{j=1}^n p((t_Z)_j) \nonumber 
	\end{eqnarray}
and 
\begin{eqnarray*}
	{\rm Tr}\,Z&=&\frac{c}{c(A',\Lambda')}\, \exp\left[\left(\bm{\mu}_{1}^{'\,T},\,\bm{\mu}_{2}^{'\,T} \right)\,M(A',\Lambda')^{-1} \left(\begin{array}{c}
		\bm{\mu}'_{1} \\ \bm{\mu}'_{2}\end{array} \right)  \right], \ \bm{\mu}=\bm{\mu}_1+i\, \bm{\mu}_2,\  \bm{\mu}_1,\bm{\mu}_2\in\mathbb{R}^{n}.
\end{eqnarray*}
\end{thm}
\begin{pf}
Follows from the detailed computations given above. 
\end{pf}

For an alternate approach on the computation of sandwiched relative $\alpha$-entropy between two gaussian states see Ref.~\cite{Seshadreesan-Lami-Wilde-2018}. 

\section*{Acknowledgement} 
This paper would have been impossible to prepare without the help of Professor A R Usha Devi. In spite of her heavy duties as Chairperson of the Physics Department, Bangalore University, Bengaluru, she has prepared the manuscript based on my handwritten computations. I also thank Mrs. Shyamala Parthasarathy for her ready support in  communicating  my handwritten papers between Delhi and Bengaluru as and when needed.

\providecommand{\MR}{\relax\ifhmode\unskip\space\fi MR }
\providecommand{\MRhref}[2]{%
	\href{http://www.ams.org/mathscinet-getitem?mr=#1}{#2}
}
\providecommand{\href}[2]{#2}

\end{document}